\documentclass[%
 reprint,
 amsmath,amssymb,
 aps,
]{revtex4-1}

\usepackage{graphicx}
\usepackage{dcolumn}
\usepackage{bm}

\begin{document}

\preprint{APS/123-QED}

\title{A decoupled scheme based on the Hermite expansion \\
to construct lattice Boltzmann models for the compressible\\
 Navier-Stokes equations with arbitrary specific heat ratio}

\author{Kainan Hu}
 \altaffiliation[Also at ]{University of Chinese Academy Sciences,Beijing ,China}
\author{Hongwu Zhang}%
 \email{Corresponding author : zhw@iet.cn}
\affiliation{%
Industrial Gas Turbine Laboratory, Institute of Engineering Thermophysics,\\
 Chinese Academy of Sciences,Beijing, China
}%

\author{Shaojuan Geng}
\affiliation{
Industrial Gas Turbine Laboratory, Institute of Engineering Thermophysics,\\
 Chinese Academy of Sciences, Beijing, China
}%



\date{\today}

\begin{abstract}
A decoupled scheme based on the Hermite expansion to construct lattice Boltzmann models  for the compressible Navier-Stokes equations  with  arbitrary specific heat ratio is proposed.
The local equilibrium distribution function including the rotational velocity of particle  is decoupled into two parts, i.e. the local equilibrium distribution function of the translational velocity of particle and that of the rotational velocity of particle.
From these two local equilibrium functions, two lattice Boltzmann models are derived  via the Hermite expansion, namely one is in relation to the translational velocity and the other is connected with the rotational velocity.
Accordingly, the distribution function is also decoupled. After this, the evolution equation is decoupled into the evolution equation of the translational velocity and that of the rotational velocity.
The two evolution equations evolve separately.
The lattice Boltzmann models used in  the scheme proposed by this work are constructed via the Hermite expansion, so it is easy to construct new schemes of  higher-order accuracy.
To validate the proposed scheme, a one dimensional shock tube  simulation is performed. The numerical results agree with  the analytical solutions very well.\\



\begin{description}
\item[PACS numbers]
47.11.-j, 47.10.-g, 47.40.-x
\end{description}
\end{abstract}

\maketitle


\section{\label{sec:sec1}Introduction}
The lattice Boltzmann method(LBM) has been successfully applied to isothermal fluids\cite{chen1992lattice,chen1998lattice}. However, when it is applied to thermal fluids, the LBM encounters some difficulties. One of them is that the specific heat ratio $\gamma$
in the macroscopic equations
derived from the Bhatnager-Gross-Krook(BGK) equation
via the Chapman-Enskog expansion
is fixed, in other words, the specific heat ratio $\gamma$ is not realistic. Several  lattice Boltzmann(LB) schemes  with   flexible specific heat ratio have been
proposed\cite{shi2001finite,kataoka2004lattice,watari2007finite,tsutahara2008new}.
These LB schemes are derived in a similar way. The discrete velocities and the local equilibrium distribution function are determined by a set of constraints which makes sure  the macroscopic equations match the thermohydrodynamic equations with certain accuracy.
Since 2006, a new way to construct LB models has been developed\cite{philippi2006continuous,philippi2015high,shan2006kinetic,mattila2014high,
shim2013multidimensional,shim2013univariate,ansumali2003minimal,chikatamarla2006entropy,chikatamarla2009lattices,yudistiawan2010higher}.
Contrary to the previous way, the new way derives the  discrete velocities and the equilibrium distribution function
via the Hermite quadrature and the Hermite expansion. The LB models constructed by the new way are more stable than the LB models constructed by the previous way and it is easy to construct LB models of higher order.
In this work, we apply the new way to   constructing  LB schemes for the compressible Navier-Stokes equations with  flexible specific heat ratio.
The local equilibrium  distribution function including the rotational velocity of particle is decoupled into two parts --- one is in relation to the translational velocity and the other is  connected with the rotational velocity.
The distribution function is also decoupled into two parts accordingly.
Two LB models are derived via the Hermite expansion.
One is for the distribution function of the translational velocity and the other is for that of the rotational velocity.  After this, we  decouple the evolution equation  into  the evolution equation of the translational velocity and that of the rotational velocity. The two evolution equations evolve separately. The decoupled scheme given above is validated by a shock tube simulation. The results of simulation agree with the analytical solutions very well.

\section{\label{sec:sec2}Decoupling the local equilibrium distribution
function including the rotational velocity of particle}
We begin with the local equilibrium distribution function. The origin that the specific heat ratio is fixed is that  gases are supposed to be monatomic, so there is only the translational free degree and  the rotational  free degree is limited. To describe  diatomic gases or polyatomic gases, the rotational velocity of particle should be introduced. 
The local equilibrium distribution function including the rotational velocity of particle is\cite{watari2007finite,xu2014modeling}
\begin{align}
f^{eq}(\bm{\xi},\eta) = & \rho \frac{1}{(2\pi R_g T)^{\frac{D}{2}}}\frac{1}{(2 n \pi R_g T)^{\frac{1}{2}}}\notag\\ &\times \exp{\Big[-\frac{(\bm{\xi}-\bm{u})^2}{2 R_g T} - \frac{\eta^2}{2 n R_g T} \Big]},
\end{align}
where $\rho$ is the density, $T$ is the absolute temperature, $\bm{u}$ is the macroscopic velocity, $\bm{\xi}$ is the translational velocity of  particle, $\eta$ is the rotational velocity of  particle, $n$ is the free degree of the rotational velocity of particle, $D$ is the dimension and $R_g$ is the universal gas constant.

The dimensionless local equilibrium distribution function is
\begin{equation}\label{Eq:DDF}
\tilde f^{eq}(\tilde{\bm{\xi}},\tilde{\eta}) = \frac{\tilde \rho}{(2 \pi \tilde \theta)^\frac{D}{2}}
 \frac{1}{(2 \pi \tilde \theta)^\frac{1}{2}}
\exp  \Big(-\frac{|\tilde {\bm{\xi}}- \tilde {\bm{u}}|^2}{2 \tilde{\theta}}\Big)
                  \exp \Big(-\frac{\tilde \eta^2}{2  \tilde{\theta}}\Big),
\end{equation}
where
$\tilde f^{eq}\!=\!f^{eq} \theta_0^{N/2} (n \theta_0 )^{1/2} $,
$\theta\!=\!R_g T$,
$\tilde \rho\!=\!\rho / \rho_0 $,
$\tilde{\bm{\xi}}\!=\!\bm{\xi}/\sqrt{\theta_0}$,
$\tilde{\eta}\!=\!\eta/n\sqrt{\theta_0}$,
$\tilde{\bm{u}}\!=\!\bm{u}/\sqrt{\theta_0}$,
$\tilde \theta\!=\!\tilde T =\theta / \theta_0$,and $\theta_0 = R_g T_0$.

Omitting the tildes on $\rho$, $\bm{\xi}$, $\eta$, $\bm{u}$, $\theta (T)$, we  simplify  Formula (\ref{Eq:DDF})
\begin{equation}
 f^{eq}(\bm{\xi},\eta) = \frac{\rho}{(2 \pi  \theta)^\frac{D}{2}}
 \frac{1}{(2 \pi  \theta)^\frac{1}{2}}
\exp \Big(-\frac{| {\bm{\xi}}-  \bm{u}|^2}{2 \theta}\Big)
\exp \Big(-\frac{ \eta^2}{2  \theta}\Big).
\end{equation}

The dimensionless local equilibrium distribution function can be decoupled into the local equilibrium distribution function of $\bm{\xi}$ and that of $\eta$
\begin{equation}\label{Eq:DF}
 f^{eq}(\bm{\xi},\eta) =  g^{eq}(\bm{\xi}) h^{eq}(\eta),
\end{equation}
where
$$ g^{eq}(\bm{\xi}) = \frac{\rho}{(2 \pi  \theta)^\frac{D}{2}}
 \exp \Big(-\frac{| {\bm{\xi}}-  \bm{u}|^2}{2 \theta}\Big),$$
$$ h^{eq}(\eta) =
\frac{1}{(2  \pi  \theta)^\frac{1}{2}}
\exp \Big(-\frac{ \eta^2}{2  \theta}\Big).$$

Taking the moment integrals of $f^{eq}(\bm{\xi})$, we obtain
\begin{subequations}
\begin{align}
\rho = &\int g^{eq}(\bm{\xi}) d \bm{\xi} \label{Eq:FeqRhoXi}\\
\rho \bm{u}= &\int g^{eq}(\bm{\xi}) \bm{\xi} d \bm{\xi} \label{Eq:FeqUXi}\\
\rho (e_t+\frac{1}{2}u^2) =& \int g^{eq}(\bm{\xi}) \frac{1}{2}\xi^2 d \bm{\xi} \label{Eq:FeqEXi}
\end{align}
\end{subequations}
where $e_t=\displaystyle\frac{D}{2}T$ is the translational internal energy.

Taking the moment integrals of $f^{eq}(\eta)$, we get
\begin{subequations}
\begin{align}
1 = &\int h^{eq}(\eta) d \eta \label{Eq:FeqRhoEta}\\
e_r =& \int h^{eq}(\eta) \frac{n}{2}\eta^2 d \eta \label{Eq:FeqEEta}
\end{align}
\end{subequations}
where $e_r = \displaystyle \frac{n}{2}T=\displaystyle\frac{n}{D}e_t $ is the rotational internal energy.

Taking the moment integrals of the local equilibrium distribution function$f^{eq}(\bm{\xi},\eta)$, we obtain
\begin{subequations}
\begin{align}
\rho = &\int\!\!\!\int f^{eq}(\bm{\xi},\eta) d \bm{\xi} d\eta\label{Eq:FeqRho}\\
\rho \bm{u}= &\int\!\!\!\int f^{eq}(\bm{\xi},\eta) \bm{\xi} d \bm{\xi} d\eta\label{Eq:FeqU}\\
\rho (E +\frac{1}{2}u^2) = &\int\!\!\!\int f^{eq}(\bm{\xi},\eta) (\frac{1}{2}\xi^2 + \frac{n}{2}\eta^2) d \bm{\xi} d\eta\label{Eq:FeqE}
\end{align}
\end{subequations}
where $ E=\displaystyle\frac{D+n}{2}T =\displaystyle \frac{D+n}{D}e_t$ is the  internal energy. It is the sum of the translational energy $e_t$ and the rotational energy $e_r$.

According to the kinetic theory, we get
\begin{subequations}\label{Eq:FRhoUE}
\begin{align}
\rho = &\int\!\!\!\int f(\bm{\xi},\eta) d \bm{\xi} d\eta\label{Eq:FeqRhoX}\\
\rho \bm{u}= &\int\!\!\!\int f(\bm{\xi},\eta) \bm{\xi} d \bm{\xi} d\eta\label{Eq:FeqUX}\\
\rho (E +\frac{1}{2}u^2) = &\int\!\!\!\int f(\bm{\xi},\eta) (\frac{1}{2}\xi^2 + \frac{n}{2}\eta^2) d \bm{\xi} d\eta\label{Eq:FeqEX}
\end{align}
\end{subequations}

Now we assume the translational velocity of particle $\bm{\xi}$ is independent of the rotational velocity of particle, so the distribution function $f(\bm{\xi},\eta)$ can  be decoupled into $g(\bm{\xi})$ and $h(\eta)$
\begin{equation}\label{Eq:FXiEta}
f(\bm{\xi},\eta) =  g(\bm{\xi}) h(\eta),
\end{equation}
where we define
\begin{subequations}\label{Eq:DefineXiEta}
\begin{align}
f(\bm{\xi},\eta) = \rho b(\bm{\xi},&\eta) = \rho b_1(\bm{\xi})h(\eta)\label{Eq:DefineXiEtaf},  \\
g(\bm{\xi}) =&  \int f(\bm{\xi},\eta) d \eta\label{Eq:DefineXiEtag},\\
h(\eta) = &\int b(\bm{\xi},\eta) d \bm{\xi}\label{Eq:DefineXiEtah}.
\end{align}
\end{subequations}
Section(\ref{sec:sec4}) and Appendix will  discuss the reasonableness of this assumption.

It should be noticed that $b(\bm{\xi},\eta)$ is a joint probability distribution, $b_1(\bm{\xi})$ and $h(\eta)$  are marginal probability distributions, $g(\bm{\xi})$ is the product of $\rho$ and a marginal probability distribution.

According to  Formula(\ref{Eq:FRhoUE}) and (\ref{Eq:DefineXiEta}),
the moments of the distribution function $g(\bm{\xi})$ are
\begin{subequations}\label{Eq:FFeqRhoUEnergy}
\begin{align}
\int g(\bm{\xi}) d \bm{\xi}          =& \int\!\!\!\int  f(\bm{\xi},\eta) d \bm{\xi} d \eta =\rho, \label{Eq:FFeqRhoXi}\\
\int g(\bm{\xi}) \bm{\xi} d \bm{\xi} =& \int\!\!\!\int  f(\bm{\xi},\eta) \bm{\xi} d \bm{\xi} d\eta = \rho \bm{u},  \label{Eq:FFeqUXi}\\
\int g(\bm{\xi}) \frac{1}{2}\xi^2 d \bm{\xi}
                          =& \int\!\!\!\int f(\bm{\xi},\eta) \frac{1}{2}\xi^2 d \bm{\xi}d\eta= \rho (e_t+\frac{1}{2}u^2).\label{Eq:FFeqEXi}
\end{align}
\end{subequations}

Similar to Formula(\ref{Eq:FFeqRhoUEnergy}),  the moments of the distribution function $h(\eta)$ are
\begin{subequations}\label{Eq:FFeqRhoUEnergyEta}
\begin{align}
\int h(\eta) d \eta                   =& \int\!\!\!\int b(\bm{\xi},\eta) d \bm{\xi} d \eta = 1 \label{Eq:FFeqRhoEta}\\
\int h(\eta) \frac{n}{2}\eta^2 d \eta =& \int\!\!\!\int b(\bm{\xi},\eta) \frac{n}{2}\eta^2 d \bm{\xi} d \eta = e_r\label{Eq:FFeqEEta}
\end{align}
\end{subequations}

\section{\label{sec:sec4}Decoupling the evolution equation}
We have decoupled the equilibrium distribution function $f^{eq}(\bm{\xi},\eta)$ and the distribution function
$f(\bm{\xi},\eta)$ in Section \ref{sec:sec2} 
\begin{align}
f^{eq}(\bm{\xi},\eta) =&  g^{eq}(\bm{\xi}) h^{eq}(\eta),\notag\\
 f(\bm{\xi},\eta) =&  g(\bm{\xi}) h(\eta). \notag
\end{align}

After decoupling $f^{eq}(\bm{\xi},\eta)$ and $f(\bm{\xi},\eta)$, we can decouple the evolution equation of $f(\bm{\xi},\eta)$. The evolution equation of $f(\bm{\xi},\eta)$ can be expressed as
\begin{equation}\label{Eq:EvolF}
\frac{\partial f(\bm{\xi},\eta)}{\partial t} + \bm{\xi} \cdot \nabla f(\bm{\xi},\eta) = -\frac{1}{\tau}[f(\bm{\xi},\eta)-f^{eq}(\bm{\xi},\eta)].
\end{equation}
where $\tau$ is the relaxation time.

Integrating Formula(\ref{Eq:EvolF}) on $\eta$, and substituting Formula(\ref{Eq:DefineXiEtag}) into it, we obtain the evolution equation of $g(\bm{\xi})$
\begin{equation}\label{Eq:DisEvolIC}
 \frac{\partial g(\bm{\xi}) }{\partial t} + \bm{\xi} \cdot \nabla g(\bm{\xi}) = -\frac{1}{\tau}[g(\bm{\xi}) - g(\bm{\xi})^{eq}].
\end{equation}

Substituting Formula(\ref{Eq:DF}) and Formula(\ref{Eq:FXiEta}) into Formula(\ref{Eq:EvolF}), we obtain
\begin{equation}\label{Eq:EvolFDis}
\frac{\partial g(\bm{\xi})h(\eta)}{\partial t} + \bm{\xi} \cdot \nabla g(\bm{\xi})h(\eta) = -\frac{1}{\tau}[g(\bm{\xi})h(\eta)-g^{eq}(\bm{\xi}) h^{eq}(\eta)].
\end{equation}

Expanding Formula(\ref{Eq:EvolFDis}) and simplifying it, we obtain
\begin{align}\label{Eq:EvolFDisI}
h(\eta)[\frac{\partial g(\bm{\xi})}{\partial t}+ &\bm{\xi}\cdot \nabla g(\bm{\xi})]+
g(\bm{\xi})[\frac{\partial h(\eta)}{\partial t}+  \bm{\xi} \cdot \nabla h(\eta)] \notag\\
& = -\frac{1}{\tau}[g(\bm{\xi})h(\eta)-g^{eq}(\bm{\xi}) h^{eq}(\eta)].
\end{align}

Substituting Formula(\ref{Eq:DisEvolIC}) into (\ref{Eq:EvolFDisI}),
integrating on $\bm{\xi}$ and simplifying it, we obtain the evolution equation of $h(\eta)$
\begin{equation}\label{Eq:EvoEtaJC}
\frac{\partial h(\eta)  }{\partial t} + \bm{u} \cdot \nabla h(\eta) = -\frac{1}{\tau} [h(\eta)  - h(\eta)^{eq}  ].
\end{equation}

Discretizing the  evolution equation of $g(\bm{\xi})$ and $h(\eta)$ in the discrete velocity space, we obtain the discrete evolution equations of $g_i$ and $h_j$
\begin{subequations}
\begin{align}
\frac{\partial g_i }{\partial t} + \bm{\xi}_i \cdot \nabla g_i = -\frac{1}{\tau}(g_i - g_i^{eq})\label{Eq:DisEvolI},\\
\frac{\partial h_j  }{\partial t} + \bm{u} \cdot \nabla h_j = -\frac{1}{\tau} (h_j  - h_j^{eq}  )\label{Eq:EvoEtaJ},
\end{align}
\end{subequations}
Where $g_i$ and $h_j$ are the discrete form of $g(\bm{\xi})$ and $h(\eta)$.
Formula(\ref{Eq:DisEvolI}) is the evolution equation of the discrete translational velocity $\bm{\xi}_i$ and formula(\ref{Eq:EvoEtaJ}) is the evolution equation of the discrete rotational velocity $\eta_j$.
It should be noticed Formula(\ref{Eq:DisEvolI}) is independent of $\eta_j$
and formula(\ref{Eq:EvoEtaJ}) is indirect connected to 
Formula(\ref{Eq:DisEvolI}) via the macroscopic velocity $\bm{u}$.

From the two  evolution equations of $g_i$ and $h_j$
i.e. Formula(\ref{Eq:DisEvolI}) and (\ref{Eq:EvoEtaJ}), the Navier-Stokes equations with  flexible specific heat ratio
via the Chapman-Enskog expansion can be derived,
\begin{subequations}
\begin{align}
\frac{\partial}{\partial t} \rho &+  \nabla \cdot \rho \bm{u} = 0,\\
\frac{\partial}{\partial t}\rho &\bm{u} +   \nabla \cdot (\rho \bm{u}\bm{u} +  P \bm{\delta})\notag\\
= & \nabla \cdot \mu [(\nabla \bm{u} + \bm{u}\nabla)-\frac{2}{D+n}\nabla \cdot \bm{u} \bm{\delta}],\\
\frac{\partial}{\partial t }\rho & (E +\frac{1}{2}u^2) +\nabla \cdot  \rho \bm{u}(E + \frac{1}{2}u^2+ \frac{P}{\rho}) \notag \\
=&\nabla \cdot  \mu \bm{u}(\nabla \bm{u} + \bm{u} \nabla - \frac{2}{D}\nabla \cdot \bm{u}\delta) + \nabla \cdot \kappa\nabla E
\end{align}
\end{subequations}
where $P\!=\!\displaystyle \frac{2}{D}\rho e_t$ is the pressure,
$\mu\!=\!\displaystyle \frac{2}{D}\rho e_t \tau $ is the dynamic viscosity coefficient,
$\kappa\!=\!\displaystyle \frac{2(D+n+2)}{D(D+n)}\rho e_t \tau$ is the heat conductivity,
and the specific heat ratio $\gamma$ is defined as
\begin{equation}
\gamma=\frac{D+n+2}{D+n}.
\end{equation}

The derivation shows 
that it is reasonable to assume   the distribution function $f(\bm{\xi},\eta)$ can be decoupled into  $g(\bm{\xi})$ and $h(\eta)$.

The appendix will give the derivation in details.

\section{\label{sec:sec3}LB models for the translational velocity and the rotational velocity }
In this section, we  derive  LB models from $g^{eq}(\bm{\xi})$ and $h^{eq}(\eta)$ respectively via the Hermite expansion. The process of deriving LB models via the Hermite expansion  has been discussed intensively by X.Shan\cite{shan2006kinetic,shan2010general}, C.Philippi\cite{philippi2006continuous,philippi2015high,mattila2014high}
and JW.Shim\cite{shim2013multidimensional,shim2013univariate}.
In this work, we only discuss  two-dimensional fluids. The case of three dimension is similar. Employing the Hermite expansion,  we  construct a two-dimensional LB model of  fourth-order accuracy, i.e. D2Q37, from $g^{eq}(\bm{\xi})$. The discrete particle velocities $\bm{\xi}_i$ and the weights $\omega_i$ of D2Q37 are showed in Table(\ref{tab:D2Q37}). The  discrete equilibrium distribution function $g^{eq}_i(\bm{\xi})$ of  D2Q37 is\begin{equation}
g^{eq}_i(\bm{\xi}) = \omega_i \rho \sum_{k=0}^{4}\frac{1}{k!}\bm{a}^{(k)}\cdot\bm{H}^{(k)},
\end{equation}
where \\
\begin{align}
 \bm{a}^{(0)}\cdot\bm{H}^{(0)}= & 1,\notag\\
 \bm{a}^{(1)}\cdot\bm{H}^{(1)}= & \bm{\xi} \cdot \bm{u},\notag\\
 \bm{a}^{(2)}\cdot\bm{H}^{(2)}= & (\bm{\xi} \cdot \bm{u})^2 + (\theta -1) (\theta^2 -D) - u^2,\notag\\
 \bm{a}^{(3)}\cdot\bm{H}^{(3)}= & (\bm{\xi} \cdot \bm{u})[(\bm{\xi} \cdot \bm{u})^2 - 3u^2 \notag\\
                                & + 3(\theta -1)( u^2 - D -2)],\notag\\
 \bm{a}^{(4)}\cdot\bm{H}^{(4)}= & (\bm{\xi} \cdot \bm{u})^4 - 6(\bm{\xi} \cdot \bm{u})^2 u^2 + 3u^4 \notag\\
                                & + 6(\theta -1)[(\bm{\xi} \cdot \bm{u})^2(u^2 - D - 4) \notag\\
                                & + (D + 2 - u^2 )\xi^2] \notag\\
                                & + 3(\theta -1)^2[u^4 - 2(D+2)u^2+D(D+2)], \notag
\end{align}
and $D\!=\!2$.

\begin{table}[!htbp]
\caption{  Discrete velocities and weights  of D2Q37.
Perm denotes permutation and $k$ denotes the number of
 discrete velocities included in each group.Scaling factor is $r =\! 1.1969797752$.}
 \begin{ruledtabular}
\begin{tabular}{ p{30pt} p{60pt}  p{80pt}   }

 $k$        & $\bm{\xi}_i$   & $\omega_i$      \\
\hline
  $1 $ & $(0,0)$       & $2.03916918e\!-\!1$      \\
  $4 $ & $Perm(r,0)$   & $1.27544846e\!-\!1$      \\
  $4 $ & $Perm(r,r)$   & $4.37537182e\!-\!2$      \\
  $4 $ & $Perm(2r,0)$  & $8.13659044e\!-\!3$      \\
  $4 $ & $Perm(2r,r)$  & $9.40079914e\!-\!3$      \\
  $4 $ & $Perm(3r,0)$  & $6.95051049e\!-\!4$      \\
  $4 $ & $Perm(3r,r)$ & $3.04298494e\!-\!5$      \\
  $4 $ & $Perm(3r,3r)$ & $2.81093762e\!-\!5$      \\
\end{tabular}
\end{ruledtabular}
\label{tab:D2Q37}
\end{table}

It should be noticed that the LB model given above is different from the LB model given by\cite{philippi2006continuous}.

In a similar way, a one-dimensional  LB model of  fourth-order accuarcy can be derived from $h^{eq}(\eta)$. Here, we adopt the D1Q7 model proposed by JW.Shim\cite{shim2013univariate}.The discrete velocities $\eta_j$ and the weights  $\omega_j$ are shown in Table(II).
\begin{table}[!htbp]
\label{tab:DxQx}
\caption{Discrete velocities and weights  of D1Q7.
 $k$ denotes the number of
 discrete velocities included in each group. Scaling factor is $r =\! 1.1969797752$.}
\begin{ruledtabular}
\begin{tabular}{ p{30pt} p{50pt}  p{80pt}   }
 $k$        & $\eta_j$   & $\omega_j$      \\
\hline
  $1 $ & $0    $       & $4.766698882e\!-\!1$      \\
  $2 $ & $\pm r$       & $2.339147370e\!-\!1$      \\
  $2 $ & $\pm 2r$      & $2.693818936e\!-\!2$      \\
  $2 $ & $\pm 3r$      & $8.121295330e\!-\!4$      \\
\end{tabular}
\end{ruledtabular}
\end{table}
The discrete equilibrium distribution function is
\begin{equation}
h^{eq}_j(\eta) = \omega_j  \sum_{j=0}^{4}\frac{1}{j!}\bm{a}^{(j)}\cdot\bm{H}^{(j)},
\end{equation}
where
\begin{align}
 \bm{a}^{(0)}\cdot\bm{H}^{(0)} = & 1,\notag\\
 \bm{a}^{(1)}\cdot\bm{H}^{(1)} = & 0,\notag\\
 \bm{a}^{(2)}\cdot\bm{H}^{(2)} = &  (\theta -1) (\theta^2 -D),\notag\\
 \bm{a}^{(3)}\cdot\bm{H}^{(3)} = & 0,\notag\\
 \bm{a}^{(4)}\cdot\bm{H}^{(4)} = & 3(\theta -1)^2[\eta^4 - 2(D+2)\eta^2+D(D+2)]\notag,
\end{align}
and $D\!=\!1$.

D2Q37 and D1Q7 are both models of  fourth-order accuracy, so the  scheme given above is of fourth-order accuracy. In this way, the higher-order of accuracy can be achieved easily.

We can also construct or adopt other LB models. But it should be noticed that the scaling factors $r$ of LB models derived form $g^{eq}(\bm{\xi})$ and  $h^{eq}(\eta)$ should be equal or else interpolation is necessary.

\section{Calculation procedure}
In this section, we first discretize the evolution equations of $g(\bm{\xi})$ and $h(\eta)$ in time and space, then we give the computational algorithm.
\subsection{Discretized the evolution equation \\ in space and time}
Now we  discretize  the discrete evolution equations of $\bm{\xi}_i$ and $\eta_j$ in time and space. The first order difference is employed for the time discretization and the convection term is performed by the third order upwind scheme. The discretized form of Formula(\ref{Eq:DisEvolI}) is
\begin{equation}\label{Eq:DisEvolSpaTiI}
g_i(\bm{x},t + \Delta t) = g_i(\bm{x},t) - \Delta t \bm{\xi}_i \cdot \nabla g_i -  \frac{\Delta t}{\tau}[g_i(\bm{x},t) -g_i^{eq}(\bm{x},t)],
\end{equation}
where $\Delta t$ is the time increment,
and the convection term along the coordinate $x$   is
\[
\begin{split}
\xi_{ix} \frac{\partial g_{i}}{\partial x} = &\frac{1}{2}\frac{\xi_{i,x}+|\xi_{i,x}|}{6\Delta x}[g_i(x\!-\!2\Delta x,y)- 6g_i(x\!-\!\Delta x,y)\\
&+ 3g_i(x,y) + 2g_i(x\!+\!\Delta x,y)] \\
&+\frac{1}{2}\frac{\xi_{i,x}-|\xi_{i,x}|}{6\Delta x}[-g_i(x\!+\!2\Delta x,y) + 6g_i(x\!+\!\Delta x,y)\\
&- 3g_i(x,y) - 2g_i(x\!-\!\Delta x,y)],
\end{split}
\]
and $\Delta x$ is the space increment. The convection term along the $y$ coordinate is similar.
In a similar way,  the discretized form of the discrete evolution equation of $\eta_j$, i.e. Formula(\ref{Eq:EvoEtaJ}), is
\begin{equation}\label{Eq:DisEvolSpaTiJ}
h_j(\bm{x},t + \Delta t) = h_j(\bm{x},t) - \Delta t \bm{u}\cdot \nabla h_j -  \frac{\Delta t}{\tau}[h_j(\bm{x},t) -h_j^{eq}(\bm{x},t)],
\end{equation}
where the convection term  is similar with that of Formula(\ref{Eq:DisEvolI})
\[
\begin{split}
u_x \frac{\partial h_{j}}{\partial x}=& \frac{1}{2}\frac{u_x+|u_x|}{6\Delta x}[h_j(x\!-\!2\Delta x,y)- 6h_j(x\!-\!\Delta x,y) \\
&+ 3h_j(x,y) + 2h_j(x\!+\!\Delta x,y)] \\
&+\frac{1}{2}\frac{u_x-|u_x|}{6\Delta x}[-h_j(x\!+\!2\Delta x,y) + 6h_j(x\!+\!\Delta x,y)\\
&- 3h_j(x,y) - 2h_j(x\!-\!\Delta x,y)].
\end{split}
\]
The convection term along the $y$ coordinate is similar.
\subsection{Computational algorithm}
The computational algorithm is as fellow :

(1) Update  $g_i$  by Formula(\ref{Eq:DisEvolSpaTiI});

(2) Update  $h_j$  by Formula(\ref{Eq:DisEvolSpaTiJ});

(3) Calculate the density $\rho$, the macroscopic velocity $\bm{u}$ and the  translational internal  energy $e_t$
\begin{subequations}
\begin{align}
\rho =& \sum_i g_i \\
\rho \bm{u} =& \sum_i g_i \bm{\xi}_i \\
\frac{1}{2}\rho u^2 + \rho e_t =& \sum_i g_i \frac{1}{2}\xi^2_i \label{Eq:TransEnergy}
\end{align}
\end{subequations}
where the translational internal energy $e_t$ is
\begin{equation}\label{Eq:e}
  e_t = \frac{D}{2}T .
\end{equation}

The pressure is defined as $P = \displaystyle\frac{2}{D} \rho e_t $;

(4) Calculate the  rotational internal energy $e_r$  
\begin{equation}\label{Eq:RotaEnergy}
e_r = \sum_j \frac{n}{2} h_j \eta^2_j,
\end{equation}
and combine Formula(\ref{Eq:RotaEnergy}) with(\ref{Eq:TransEnergy}), then we obtain
\begin{equation}\label{Eq:Total}
\frac{1}{2}\rho u^2 + \rho E =\sum_i g_i \frac{1}{2}\xi^2_i + \rho \sum_j h_j \frac{n}{2} \eta^2_j .
\end{equation}
As defined above, the internal energy $E$ is the sum of the translational internal energy $e_t$ and the rotational internal energy
$e_r$
\begin{equation}\label{Eq:InernalEnergy}
E = e_t + e_r = \frac{D+n}{D}e_t.
\end{equation}
Substituting Formula(\ref{Eq:InernalEnergy}) into (\ref{Eq:Total}) we obtain the internal energy $E$
\begin{equation}\label{Eq:E}
E =\frac{\displaystyle\sum_i g_i \frac{1}{2}\xi^2_i + \rho\sum_j \frac{n}{2} h_j \eta^2_j - \rho\frac{1}{2}u^2}{\rho} .
\end{equation}
Substituting Formula(\ref{Eq:e}) and (\ref{Eq:InernalEnergy}) into (\ref{Eq:E}) we obtain the absolute temperature $T$
\begin{equation}\label{Eq:T}
  T = \frac{2}{D+n}\frac{\displaystyle\sum_i g_i \frac{1}{2}\xi^2_i + \rho\sum_j \frac{n}{2} h_j \eta^2_j - \rho\frac{1}{2}u^2}{\rho}.
\end{equation}

(5) Implement the boundary conditions.  


\section{Numerical validation}
In this section, we apply the decoupled scheme given above to simulating a shock tube. The grid is $X \times Y \!=\!  1000 \times 16 $. The initial condition of the left tube is ${\rho \!=\! 4,T\!=\!1,\bm{u}\!=\!0}$ and that of the right tube is ${\rho \!=\! 1,T\!=\!1,\bm{u}\!=\!0}$. The specific heat ratio is $\gamma  \!=\!1.4$, the rotational  free degree is $n\!=\!3$ and the relaxation time is $\tau\!=\!2/3$.
All of these macroscopic variables are dimensionless.
The periodic boundary condition is employed for the up and down boundaries and the open boundary condition is employed for the left and right boundaries.

\begin{figure*}[!hbt]
\includegraphics{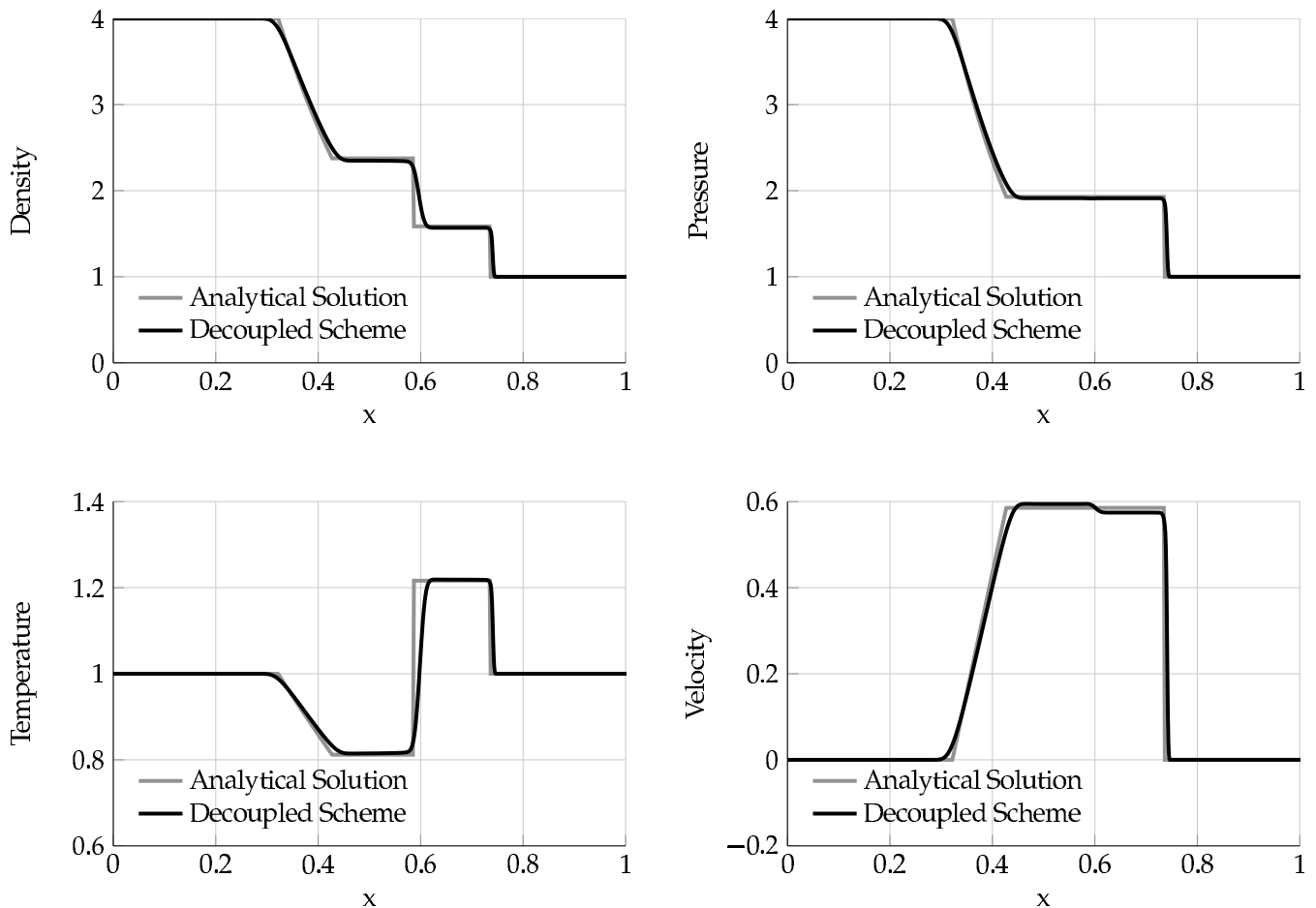}
\caption{The black lines are the simulation results and the gray lines are the analytical resolutions  .These are the results at the 180th step, i.e.  the time $t\!=\!0.1504$. The specific heat ratio is $\gamma\! = \! 1.4$ and the relaxation time is $\tau\!=\! 2/3$. The initial condition of the left side is ${\rho \!=\! 4,T\!=\!1,\bm{u}\!=\!0}$ and that of the right side is ${\rho \!=\! 1,T\!=\!1,\bm{u}\!=\!0}$.}
\label{Fig:HeatRatio}
\end{figure*}

Fig(\ref{Fig:HeatRatio}) gives the results of simulation employing the decoupled scheme given above at   $step = 180$, i.e.   time  $$t = \displaystyle{\frac{step}{ X \times r}}= \displaystyle{\frac{180}{1000\times 1.1969797752}} = 0.1504.$$ The analytical solutions~\cite{SOD19781} at the same time are also given.
The black lines show the simulation results and the gray lines show the analytical resolutions. It  can be seen from Fig(\ref{Fig:HeatRatio}) that the simulation results agree with the analytical resolutions very well.

Tab(\ref{tab:RelativeError}) gives the relative error of density $\rho$, pressure $p$, absolute temperature $T$ and the velocity $u$. The relative error is defined as
\begin{equation}\label{Eq:Error}
 Error = \frac{ \displaystyle\sum_i |x_{nume, i}-x_{anal, i}|}{\displaystyle\sum_i |x_{anal,i}|},
\end{equation}
where $x_{nume}$ is the numerical solutions, $x_{anal}$ is the analytical solutions and $i = X+1$. Tab(\ref{tab:RelativeError}) shows that
the maximum of relative error is $2.42\%$. This relative error is acceptable.

\begin{table}[!htbp]
\label{tab:RelativeError}
\caption{Relative error of the numerical solutions.}
\begin{ruledtabular}
\begin{tabular}{ l l l l l  }
       & $\rho$   & $p$  & $T$       & $u_x$   \\
\hline
  $Error$ & $0.0077$   & $0.0055$   &$0.0064$ &  $0.0242$ \\
\end{tabular}
\end{ruledtabular}
\end{table}
\section{Conclusion}

This work proposes a way based on the Hermite expansion to construct  LB schemes for the compressible Navier-Stokes equations with  arbitrary specific heat ratio.
The equilibrium distribution function $f^{eq}(\bm{\xi},\eta)$, the distribution function $f(\bm{\xi},\eta)$ and the evolution function are decoupled into two parts, namely one is in relation to the translational velocity $\bm{\xi}$ and the other is connected with the rotational velocity $\eta$.
The two evolution equations evolve separately.
The translational velocity $\bm{\xi}$ is discretized in a two- or three- dimensional LB model and
the rotational velocity $\eta$  is discretized in another  one-dimensional LB model.
The Hermite expansion is applied to deriving these two LB models.
The correct flexible specific heat ratio is obtained
and correct relation between the temperature $T$ and the internal energy $E$ is derived via the Chapman-Enskog expansion.
The decoupled scheme is validated by a shock tube simulation.
The simulation results  agree with the analytical resolutions  very well.

The LB models used in the decoupled scheme is same as the ones used in the schemes with fixed specific heat ratio.
It is not necessary for the decoupled scheme to construct new LB models specially.
The models with fixed specific heat ratio can applied to the decoupled scheme without any recommendation.
This is different from the existing schemes which construct new models in order to adjust the specific heat ratio.

The decoupled scheme proposed by this work can make use of the models  constructed via the Hermite expansion, so the process of constructing new schemes is simple and   higher-order  accuracy can be achieved  easily. For the same reason, the decoupled scheme is more stable than the schemes constructed by the  early way, which derives LB models via a try-error method.

\appendix*
\section{Derivation of the Navier-Stokes equations from the  evolution equations of $g(\bm{\xi})$ and $h({\eta})$ via the Chapman-Enskog expansion}
In this appendix, we derive the Navier-Stokes equations with  flexible specific heat ratio from the   evolution equations of $g(\bm{\xi})$ and $h({\eta})$ via the Chapman-Enskog expansion.

Expanding the distribution functions $g_i$ and $h_j$, the derivatives of the time $t$  and the space in terms of the Kundsen number $\epsilon$ we obtain
\begin{subequations}\label{Eq:fij}
\begin{align}
\nabla =& \epsilon \nabla_1 ,\\
\frac{\partial}{\partial t} = \epsilon \frac{\partial}{\partial t_1} &+ \epsilon^2 \frac{\partial}{\partial t_2},\label{Eq:TimeScale}\\
g_i = g_i^{(0)} +  \epsilon &g_i^{(1)}+ \epsilon^2  g_i^{(2)} ,\label{Eq:f_i}\\
h_j = h_j^{(0)} +  \epsilon &h_j^{(1)}+ \epsilon^2  h_j^{(2)} .\label{Eq:f_j}
\end{align}
\end{subequations}

Substituting Formula(\ref{Eq:f_i}) into the evolution equation of the translational velocity, i.e.  Formula(\ref{Eq:DisEvolI}) and comparing the order of $\epsilon$ we obtain
\begin{subequations}
\begin{align}
&g_i^{(0)}=g_i^{(eq)},\\
(\frac{\partial}{\partial t_1} + &\bm{\xi}_i \cdot \nabla_1)g_i^{(0)} + \frac{1}{\tau}g_i^{(1)}=0,\label{Eq:XiOrder1}\\
\frac{\partial g_i^{(0)}}{\partial t_2}+(\frac{\partial}{\partial t_1} &+ \bm{\xi}_i \cdot \nabla_1)g_i^{(1)} + \frac{1}{\tau}g_i^{(2)}=0.\label{Eq:XiOrder2}
\end{align}
\end{subequations}

Considering the discrete form of Formula(\ref{Eq:FFeqRhoUEnergy}) in the discrete velocity space
\begin{subequations}
\begin{align}
\sum_i g_i =&  \sum_i g_i^{eq} = \rho  ,\\
\sum_i g_i =&  \sum_ig_i^{eq}  = \rho \bm{u}  ,\\
\sum_i g_i \frac{1}{2}\xi_i^2 =& \sum_ig_i^{eq}\frac{1}{2}\xi_i^2 = \rho (\frac{1}{2}u^2+ e_t ) ,
\end{align}
\end{subequations}
we obtain
\begin{equation}\label{Eq:f_i_zero}
\sum_i g_i^{(n)} = 0,
\sum_i g_i^{(n)}\bm{\xi}_i=0,
\sum_i g_i^{(n)}\frac{1}{2}\xi_i^{2}=0,\quad n=1,2.
\end{equation}

Substituting Formula(\ref{Eq:f_j}) into the evolution equation of the rotational velocity,i.e.Formula(\ref{Eq:EvoEtaJ}) and comparing the order of $\epsilon$ we obtain
\begin{subequations}
\begin{align}
&h_j^{(0)}=h_j^{(eq)},\\
(\frac{\partial}{\partial t_1} + &\bm{u} \cdot \nabla_1)h_j^{(0)} + \frac{1}{\tau}h_j^{(1)}=0,\label{Eq:order1}\\
\frac{\partial h_j^{(0)}}{\partial t_2}+(\frac{\partial}{\partial t_1} &+ \bm{u} \cdot \nabla_1)h_j^{(1)} + \frac{1}{\tau}h_j^{(2)}=0.\label{Eq:order2}
\end{align}
\end{subequations}

Considering the discrete form of Formula(\ref{Eq:FFeqRhoUEnergyEta}) in the discrete velocity space
\begin{subequations}
\begin{align}
\sum_j h_j = &\sum_j h_j^{eq} = 1 ,\\
\sum_j h_j \frac{n}{2}\eta_j^2 = &\sum_j h_j^{eq}\frac{n}{2}\eta_j^2 = e_r ,
\end{align}
\end{subequations}
we obtain
\begin{equation}\label{Eq:f_j_zero}
\sum_j h_j^{(n)} = 0,
\sum_j h_j^{(n)}\frac{n}{2}\eta^2_j=0,\quad n=1,2.
\end{equation}

Some velocity moments of $g_i$ and $h_j$ will be used in the derivation of the Navier-Stokes equations and we list them as fellow
\begin{subequations}\label{Eq:Moment}
\begin{align}
\sum_i g_i^{eq} = & \rho,\\
\sum_i g_i^{eq}\bm{\xi}_i = &\rho \bm{u},\\
\sum_i g_i^{eq}\bm{\xi}_i \bm{\xi}_i  =& \rho \bm{u} \bm{u} + P\bm{\delta},\\
\sum_i g_i^{eq}\bm{\xi}_i \bm{\xi}_i \bm{\xi}_i  =& \rho \bm{u} \bm{u} \bm{u} + P\bm{u}\bm{\delta},\\
\sum_i g_i^{eq}\frac{1}{2} \xi_i^2 =& \rho(\frac{1}{2} u^2 +  e_t) ,\\
\sum_i g_i^{eq}\frac{1}{2} \xi_i^2 \bm{\xi}_i = & \rho(\frac{1}{2} u^2 + e_t)\bm{u} ,\\
\sum_i g_i^{eq}\frac{1}{2} \xi_i^2 \bm{\xi}_i\bm{\xi}_i = &P(\frac{2}{D}e_t + \frac{1}{2} u^2 +  e_t)\bm{\delta}, \notag\\
& +(2P+\frac{1}{2}\rho u^2 + \rho e_t)\bm{u}\bm{u},
\end{align}
\end{subequations}
where $P\bm{u}\bm{\delta} = P(u_r \delta_{st} + u_s \delta_{tr} + u_t \delta_{rs})$. Here, the Grad notes is used\cite{grad1949note}.
Two velocity moments of $h_j$ will be used in the following parts
\begin{subequations}\label{Eq:MomentsFj}
\begin{align}
\sum_j h_j^{eq} &= 1,\\
\sum_j h_j^{eq} \frac{n}{2}\eta^2 & = e_r.
\end{align}
\end{subequations}

\subsection{Derivation of the continuity equation}
Taking the zeroth order moment of  Formula(\ref{Eq:XiOrder1}), we obtain
\begin{equation}\label{Eq:f_i_sum}
\sum_i(\frac{\partial}{\partial t_1} + \bm{\xi}_i \cdot \nabla_1) g_i^{(0)} + \frac{1}{\tau}\sum_i g_i^{(1)}=0
\end{equation}
Substituting Formula(\ref{Eq:f_i_zero}) and (\ref{Eq:Moment}) into (\ref{Eq:f_i_sum}), the continuity equation of the first order is obtained
\begin{equation}\label{Eq:RhoOrder1}
\frac{\partial}{\partial t_1}\rho +  \nabla_1 \cdot \rho \bm{u} = 0.
\end{equation}

Taking the zeroth order moment of  Formula(\ref{Eq:XiOrder2}), we obtain
\begin{equation}\label{Eq:fiSumOrder2}
\frac{\partial \sum_i g_i^{(0)}}{\partial t_2}+\sum_i (\frac{\partial}{\partial t_1} + \bm{\xi}_i \cdot \nabla_1)g_i^{(1)} + \frac{1}{\tau}\sum_i g_i^{(2)}=0.
\end{equation}
Substituting Formula(\ref{Eq:f_i_zero}) and (\ref{Eq:Moment}) into (\ref{Eq:fiSumOrder2}), then summing on $i$ we obtain the continuity equation of the second order
\begin{equation}\label{Eq:RhoOrder2}
\frac{\partial \rho}{\partial t_2}=0.
\end{equation}
Making use of Formula(\ref{Eq:TimeScale}) and combining  the continuity equation of the first and second order, i.e. Formula(\ref{Eq:RhoOrder1}) and (\ref{Eq:RhoOrder2}), the continuity equation is obtained
\begin{equation}
\frac{\partial}{\partial t} \rho +  \nabla \cdot \rho \bm{u} = 0.
\end{equation}
\subsection{Derivation of the momentum \\ conservation equation}
Taking the first order moment of Formula(\ref{Eq:XiOrder1})
\begin{equation}
\sum_i(\frac{\partial}{\partial t_1} + \bm{\xi}_i \cdot \nabla_1) g_i^{(0)}\bm{\xi}_i + \frac{1}{\tau}\sum_i g_i^{(1)}\bm{\xi}_i=0,\\
\end{equation}
and inserting Formula(\ref{Eq:f_i_zero}) and (\ref{Eq:Moment}), we get the conservation  momentum equation of the first order
\begin{equation}\label{Eq:RhoXiOrder1}
\frac{\partial}{\partial t_1}\rho \bm{u} +   \nabla_1 \cdot (\rho \bm{u}\bm{u} + P \bm{\delta} ) = 0.\\
\end{equation}

Taking the first order moment of Formula(\ref{Eq:XiOrder2})
\begin{equation}\label{Eq:fiMomentOne}
\frac{\partial \sum_i g_i^{(0)}\bm{\xi}_i}{\partial t_2} + \sum_i(\frac{\partial}{\partial t_1} + \bm{\xi}_i \cdot \nabla_1) g_i^{(1)}\bm{\xi}_i + \frac{1}{\tau}\sum_i g_i^{(2)}\bm{\xi}_i=0,
\end{equation}
substituting Formula(\ref{Eq:XiOrder1}) into (\ref{Eq:fiMomentOne}) and simplifying, we obtain
\begin{align}\label{Eq:fiMomentTwoA}
\frac{\partial \sum_i g_i^{(0)}\bm{\xi}_i}{\partial t_2}
-\tau\nabla_1\cdot&(\frac{\partial}{\partial t_1}\sum_i \bm{\xi}_i \bm{\xi}_i g_i^{(0)} \notag\\
&+ \nabla_1\cdot \sum_i \bm{\xi}_i \bm{\xi}_i  \bm{\xi}_i g_i^{(1)} ) = 0.
\end{align}
Inserting the moments of $g_i$ i.e. Formula(\ref{Eq:Moment}), into Formula(\ref{Eq:fiMomentTwoA}) we obtain
\begin{equation}\label{Eq:fiMomentTwoBX}
\frac{\partial \rho \bm{u}}{\partial t_2}
-\tau\nabla_1\cdot[\frac{\partial}{\partial t_1}(\rho \bm{u}\bm{u}+P\bm{\delta}) + \nabla_1 \cdot \sum_i \bm{\xi}_i  \bm{\xi}_i  \bm{\xi}_i g_i^{(1)} ] = 0,
\end{equation}
where $\displaystyle\frac{\partial P}{\partial t_1}$ is a difficult point to simplify. To simplify  $\displaystyle\frac{\partial P}{\partial t_1}$, we should obtain the energy conservation equation of the first order firstly.

Multiplying $\displaystyle\frac{1}{2}\xi^2_i$ to Formula(\ref{Eq:XiOrder1}) and  summing on  $i$ we obtain
\begin{equation}\label{Eq:fiMoementTwoC}
\sum_i (\frac{\partial}{\partial t_1} + \bm{\xi}_i \cdot \nabla_1)g_i^{(0)}\frac{1}{2}\xi^2_i + \frac{1}{\tau}\sum_i g_i^{(1)}\frac{1}{2}\xi^2_i=0.
\end{equation}
Substituting the moments of $\bm{\xi}_i$ i.e. Formula(\ref{Eq:Moment}), into Formula(\ref{Eq:fiMoementTwoC}), we obtain the translational internal energy conversation equation of the first order
\begin{equation}\label{Eq:TranslationalEenegyOrderOne}
\frac{\partial}{\partial t_1}(\frac{1}{2}\rho u^2+\rho e_t)
 +  \nabla_1\cdot(\frac{1}{2}\rho u^2+\rho e_t + P)\bm{u} =0.
\end{equation}

Multiplying $\displaystyle\frac{1}{2}\eta^2$ to Formula(\ref{Eq:order1}),  summing on  $j$  and inserting the moments
of $h_j$ ,i.e. Formula(\ref{Eq:MomentsFj}), we obtain the rotational internal energy conversation equation of the first order in nonconservation form
\begin{equation}\label{Eq:EnEta1}
\frac{\partial}{\partial t_1}e_r + \bm{u} \cdot \nabla_1 e_r = 0.
\end{equation}

Multiplying $\rho$ to Formula(\ref{Eq:EnEta1}) and multiplying $e_r$ to Formula(\ref{Eq:RhoOrder1}), then adding up we obtain
\begin{equation}\label{Eq:EnEta}
\rho(\frac{\partial}{\partial t_1}e_r + \bm{u} \cdot \nabla_1 e_r) + e_r (\frac{\partial}{\partial t_1}\rho + \nabla_1 \cdot \rho \bm{u}) =0.
\end{equation}

Simplifying Formula(\ref{Eq:EnEta}) we obtain the rotational internal conversation energy equation of the first order  in conservation form
\begin{equation}\label{Eq:EnEtaEnd}
\frac{\partial}{\partial t_1 }\rho e_r + \nabla_1 \cdot \rho \bm{u} e_r = 0.
\end{equation}

Combining the translational internal energy conversation equation of the first order Formula(\ref{Eq:TranslationalEenegyOrderOne}) and the rotational internal energy conversation equation of the first order formula(\ref{Eq:EnEtaEnd}), we obtain the energy conversation equation of the first order
\begin{equation}\label{Eq:EenegyOrderOne}
\frac{\partial}{\partial t_1}(\frac{1}{2}\rho u^2+\rho E)
 +  \nabla_1\cdot(\frac{1}{2}\rho u^2+\rho E + P)\bm{u} =0.
\end{equation}
Substituting $P = \displaystyle \frac{2}{D+n}\rho E $ into Formula(\ref{Eq:EenegyOrderOne}) we obtain
\begin{equation}\label{Eq:EenegyOrderOneB}
\frac{\partial}{\partial t_1}(\frac{1}{2}\rho u^2+ \frac{D+n}{D}P)
 +  \nabla_1\cdot(\frac{1}{2}\rho u^2+ \frac{D+n+2}{D} P)\bm{u} =0.
\end{equation}
Expanding Formula(\ref{Eq:EenegyOrderOneB}), substituting (\ref{Eq:RhoOrder1}) and (\ref{Eq:RhoXiOrder1}) into it, after some algebra, we obtain
\begin{equation}\label{Eq:PartialP}
\frac{\partial P}{\partial t_1} = -\nabla_1 \cdot P\bm{u} - \frac{2}{D+n} P \nabla_1 \cdot \bm{u},
\end{equation}

Substituting Formula(\ref{Eq:PartialP}) into (\ref{Eq:fiMomentTwoBX}),
after some algebra, we obtain the momentum conversation equation of the second order
\begin{equation}\label{Eq:MomentEquationOrederTwo}
\frac{\partial}{\partial t_2}\rho\bm{u}=\nabla_1 \cdot \frac{2}{D} \rho e_t \tau [(\nabla_1 \bm{u} + \bm{u}\nabla_1)-\frac{2}{D+n}\nabla_1 \cdot \bm{u} \bm{\delta}].
\end{equation}

Combining the momentum equation of the first order and second order, we obtain the moment conversation equation
\begin{equation}\label{Eq:RhoXiOrder}
\frac{\partial}{\partial t}\rho \bm{u} +   \nabla \cdot (\rho \bm{u}\bm{u} +  P \bm{\delta})  = \nabla \cdot \mu [(\nabla \bm{u} + \bm{u}\nabla)-\frac{2}{D+n}\nabla \cdot \bm{u} \bm{\delta}],\\
\end{equation}
where $\mu =\displaystyle \frac{2}{D}\rho e_t\tau $ is the dynamic viscosity coefficient.
\subsection{Derivation of the energy conversation equation}
We have obtained the energy conversation equation of the first order, i.e. Formula(\ref{Eq:EenegyOrderOne}),
\begin{equation}
\frac{\partial}{\partial t_1}(\frac{1}{2}\rho u^2+\rho E)
 +  \nabla_1\cdot(\frac{1}{2}\rho u^2+\rho E + P)\bm{u} =0.\notag
\end{equation}
Now, we  derive the energy conversation equation of the second order.

Multiplying $\displaystyle\frac{1}{2}\xi_i^2$ to Formula(\ref{Eq:XiOrder2}), substituting (\ref{Eq:XiOrder1}) into it, and summing  on $i$, we obtain
\begin{align}\label{Eq:fiMomentTwoX}
\frac{\partial}{\partial t_2} \sum_i g_i^{(0)}\frac{1}{2}\xi_i^2
= \nabla_1 \cdot\tau(\frac{\partial}{\partial t_1}&\sum_i g_i^{(0)}\frac{1}{2}\xi^2_i \bm{\xi}_i \notag\\
+ & \nabla_1 \cdot  g_i^{(0)}\bm{\xi}_i\bm{\xi}_i \frac{1}{2}\xi^2_i)
\end{align}
Substituting the moments of $g_i$ into Formula(\ref{Eq:fiMomentTwoX}), we obtain
\begin{align}\label{Eq:fiMomentThree}
\frac{\partial}{\partial t_2}& ( \frac{1}{2}\rho u^2 + e_t)
=  \nabla_1 \cdot\tau [ \frac{\partial}{\partial t_1}(\frac{1}{2}\rho u^2 + e_t +P)\bm{u} \notag\\
+ & \nabla_1 \cdot  P(\frac{2}{D}e_t + \frac{1}{2}u^2 + e_t)\bm{\delta} + (2P + \frac{1}{2}\rho u^2 + e_t)\bm{u}\bm{u} ]
\end{align}

Inserting Formula (\ref{Eq:RhoOrder1}) , (\ref{Eq:RhoXiOrder1}), (\ref{Eq:EenegyOrderOneB}) and (\ref{Eq:PartialP}) into
Formula(\ref{Eq:fiMomentThree}), after some algebra, we get the translational internal energy conversation equation of the second order
\begin{align}\label{Eq:fiMomentTwoFour}
\frac{\partial}{\partial t_2}( \frac{1}{2}\rho & u^2 + e_t)
=\nabla_1 \cdot\tau P\bm{u}[(\nabla_1 \bm{u} +\bm{u}\nabla_1) \notag \\
 & -\frac{2}{D+n}\nabla_1 \bm{u} \bm{\delta}]
+ \nabla_1 \cdot\tau P\frac{D+2}{D}\nabla_1 e_t.
\end{align}

Multiplying $\displaystyle h_j^{(0)}$ to Formula(\ref{Eq:XiOrder2}), Multiplying $\displaystyle g_i^{(0)}$ to Formula(\ref{Eq:order2}) and adding them up
\begin{align}\label{Eq:fiRotationalEnergyOne}
\frac{\partial }{\partial t_2}g_i^{(0)}h_j^{(0)}
&+ (\frac{\partial}{ \partial t_1} + \xi \cdot \nabla_1) g_i^{(1)} h_j^{(0)}\notag\\
+ (\frac{\partial}{ \partial t_1} +& \bm{u} \cdot\nabla_1) g_i^{(0)} h_j^{(1)}
+ \frac{1}{\tau}(g_i^{(2)} h_j^{(0)} + g_i^{(0)} h_j^{(2)} ) = 0,
\end{align}
substituting Formula(\ref{Eq:XiOrder1}) into (\ref{Eq:fiRotationalEnergyOne}), we get
\begin{align}\label{Eq:fiRotationalEnergyTwo}
\frac{\partial }{\partial t_2}g_i^{(0)}h_j^{(0)}
&-  \nabla_1 \cdot \tau(\frac{\partial}{\partial t_1} \bm{\xi}_i  g_i^{(0)} h_j^{(0)} + \nabla_1 \cdot \bm{\xi}_i\bm{\xi}_i  g_i^{(0)} h_j^{(0)} )   \notag\\
+ (\frac{\partial}{ \partial t_1} +& \bm{u} \cdot\nabla_1) g_i^{(0)} h_j^{(1)}
+ \frac{1}{\tau}(g_i^{(2)} h_j^{(0)} + g_i^{(0)} h_j^{(2)} ) = 0.
\end{align}

Multiplying $\displaystyle \frac{n}{2}\eta^2$  to Formula(\ref{Eq:fiRotationalEnergyTwo}) and summing on $i$ and $j$ we obtain
\begin{align}\label{Eq:fiRotationalEnergyThree}
\frac{\partial }{\partial t_2}\rho e_r
 = \nabla_1 \cdot \tau[\frac{\partial}{\partial t_1} \rho \bm{u} e_r + \nabla_1 \cdot (\rho \bm{u}\bm{u}+P\bm{\delta})e_r].   \notag\\
\end{align}
Inserting Formula(\ref{Eq:RhoXiOrder1}) and (\ref{Eq:EnEtaEnd}) into (\ref{Eq:fiRotationalEnergyThree}), after some algebra, we obtain the rotational internal energy conversation equation of the second order
\begin{align}\label{Eq:fiRotationalEnergyFour}
\frac{\partial }{\partial t_2}\rho e_r
 = \nabla_1 \cdot \tau \rho e_t \frac{n}{D}\nabla_1 e_r.
\end{align}

Combining the rotational internal energy conversation equation of the seconde order Formula(\ref{Eq:fiRotationalEnergyFour}) with the translational internal conversation energy of the seconde order Formula(\ref{Eq:fiMomentTwoFour}), we obtain the energy conversation  equation of the second order
\begin{align}\label{Eq:EnergySecondOreder}
\frac{\partial}{\partial t_2}&(\frac{1}{2}\rho  u^2 + \rho E)
=\nabla_1 \cdot\tau P\bm{u}[(\nabla_1 \bm{u} +\bm{u}\nabla_1) \notag \\
& -\frac{2}{D+n}\nabla_1 \bm{u} \bm{\delta}]
+ \nabla_1 \cdot \frac{D+n+2}{D} \tau \rho e_t\nabla_1 e_t.
\end{align}

Combining the energy conversation equation of the second order (\ref{Eq:EnergySecondOreder}) with the energy conversation equation of the first order
Formula(\ref{Eq:EenegyOrderOne}) we obtain the energy conversation equation
\begin{align}\label{Eq:EnergyEquation}
\frac{\partial}{\partial t }\rho &(E +\frac{1}{2}u^2) + \nabla \cdot \rho \bm{u}(E + \frac{1}{2}u^2 + \frac{P}{\rho}) \notag \\
=& \nabla \cdot \mu \bm{u}(\nabla \bm{u} + \bm{u} \nabla - \frac{2}{D}\nabla \cdot \bm{u}\delta) + \nabla \cdot \kappa\nabla E
\end{align}
where
\begin{align}
P=\frac{2}{D}\rho e_t,
\mu = \frac{2}{D}\rho e_t \tau,
\kappa =\frac{2(D+n+2)}{D(D+n)}\rho e_t \tau.
\end{align}
Formula(\ref{Eq:EnergyEquation}) is the energy conversation equation with flexible specific heat ratio
and the specific heat ratio $\gamma$ is
\begin{equation}
\gamma=\frac{D+n+2}{D+n}.\notag
\end{equation}
The specific heat ratio $\gamma$ can be adjusted by changing the free degree of the rotational velocity $n$. 

\nocite{*}

\bibliography{HeatRatio}

\end{document}